# Understanding the behavioural difference of PPCA among its homologs in $C_7$ family towards recognition of DXCA

**Suvankar Ghosh[2] • Shankar Kumar Ghosh[1] • Camellia Ray[1] • Goutam Paul[1] • Pabitra Pal Choudhury[1] • Raja Banerjee[2]**

[1] Indian Statistical Institute, 203 B.T. Road, Kolkata-700108, India
[2] West Bengal University of Technology, BF-142,
Sector-1, Salt Lake, Kolkata- 700064, India

**Abstract**
Among all the proteins of Periplasmic C type cytochrome A (PPCA) family obtained from cytochrome $C_7$ found in Geobacter sulfurreducens, PPCA protein can interact with Deoxycholate (DXCA), while its other homologs do not, as observed from the crystal structures. Utilizing the concept of ‗structure-function relationship', an effort has been initiated towards understanding the driving force for recognition of DXCA exclusively by PPCA among its homologs. Further, a combinatorial analysis of the binding sequences (contiguous sequence of amino acid residues of binding locations) is performed to build graph-theoretic models, which show that PPCA differs from its homologues. Analysis of the results suggests that the underlying impetus of recognition of DXCA by PPCA is embedded in its primary sequence and 3D conformation.

## 1. Introduction

*Geobacter sulfurreducens*, one of the predominant metal and sulphur reducing bacteria (Bond and Lovley 2003), found below the surface of earth, are comma shaped gram-negative, anaerobic bacteria. This organism can act as an electron donor and participate in redox reaction (Caccavo et al. 1994). This ability can be used to increase the effectiveness of microbial fuel (Caccavo et al. 1994). *G. sulfurreducens* encodes over 100 cytochrome C and several of them have important roles in the respiration of this organism under various conditions (Shelobolina et al. 2007; Shi et al. 2007). Periplasmic C type cytochrome A (PPCA) family proteins, one of the cytochromes of the $C_7$ family found in *Geobacter sulfurreducens*, and are used for the reduction of Fe(III) (Lloyd et al. 2003). Besides this, PPCA can interact with deoxycholate (DXCA), also known as deoxycholic acid, which is a byproduct of intestinal bacteria, used in the medicinal field to emulsify fats for the absorption in the intestine (Rotunda et al. 2005). DXCA is also used in the research field as a mild detergent, for the isolation of membrane associated proteins (Burgess and Deutscher 1990).

As evident from the crystal structure that , among the PPCA family proteins found in Geobacter sulfurreducens, only PPCA can interact with Deoxycholate (DXCA), while its other homologues cannot although they have high sequence identity with PPCA (Pokkuluri et al. 2011; Pokkuluri et al. 2004). Moreover, towards interaction with DXCA, 4, 29, 37, 38, 41, 45 and 50 number residues of PPCA are utilized (Pokkuluri et al. 2011; Pokkuluri et al. 2004; Morgado et al. 2012). At this prevailing situation it would be worthy to identify the basic reason of such an amazing difference towards recognizing a single compound between the homologous proteins having high sequence similarity.

An attempt has been initiated utilizing the concept of ‗structure-function relationship' as the information regarding the secondary and tertiary structure of proteins is embedded in its primary amino acid sequence (Anfinsen 1973) and hence the functional aspect. This may provide a guideline to rationalize the driving force for recognition of DXCA exclusively by PPCA among its homologues. However, to ensure the basic motif of such an astonishing difference between the homologous proteins towards recognition of a single compound, we feel it would be admirable to go backward towards nucleotide sequence. Towards this end, a combinatorial analysis of binding sequences which consists of the nucleotide bases of binding locations as well as its homologues, have been done through building ~~a~~ graph-theoretic models At this point of argument one may raise a question that when information about protein sequence and structure is available, why the genomic sequence is used to perform the analyses. Towards our understanding, such a route map from óDNA to proteinô utilizing the combinatorial approach would address the problem from the órootô as all the biological information is primarily embedded in DNA, the blueprint of life (Alberts et al. 2002, Leavitt 2010, Slack et al. 2014), which passes to protein for its function. Such sequence/structural analysis from the origin of sequence utilizing discrete mathematical model would be able to explain the behavioral difference of PPCA within the homologues which would have an added value to the basic biology.

## 2. Materials and Methods

Primary sequence and the 3D coordinate file of PPCA protein and its homologs (PPCB, PPCC, PPCD and PPCE) have been obtained from the Protein Data Bank (http://www.rcsb.org). Corresponding PDB IDs of the respective protein are: 1OS6: PPCA; 3BXU: PPCB; 3H33: PPCC; 3H4N: PPCD and 3H34: PPCE.

### 2.1. Sequence and structural alignments

Sequence identity of each of the PPCA protein homologs with respect to PPCA is obtained by pairwise alignment of sequences using a sequence alignment tool óCLUSTAL Wô (Larkin et al. 2007).
Further, 3D-structure alignments of these proteins have been performed using óSuperPose serverô (Maiti et al. 2004) which also helps to cross check the sequence alignments, as well as gives an idea about the RMSD statistics, Different Distance Plots, and interactive images of the superimposed structures.

### 2.2. Conformational analysis of PPCA family

Conformational analysis of PPCA protein homologs has been pursued using DSSP program (Kabsch and Sander 1983) on the respective PDB entries which gives an overview of the detailed secondary structure that includes main chain torsion angles (ϕ, ɣ) as well as virtual bond angle (ə), virtual torsion angle (Ŭ), nature of hydrogen bond between atoms.

### 2.3. Graph Theoretic Terminologies

In this section we have introduced some graph theoretic terminologies from (Deo N. 2003) which we have used in our work.

- A *graph* G=(V, E) consists of a set of objects V= $\{v_1, v_2, é, v_n\}$ called *vertices* and another set E = $\{e_1, e_2, é, e_m\}$ whose elements are called *edges*, such that each *edge* $e_k$ , k=1, 2, 3,é *m* is identified as a pair of vertices $(v_i,v_j)$ , $i, j \in \{1,2,3,é\ n\}$ where $v_i$ and $v_j$ are called *end vertices*. G is called *directed* graph if $(v_i,v_j)$ is an *ordered pair* whereas *undirected* if they are *unordered pair*, written as $\{v_i,v_j\}$.

For example, the directed graph shown in Fig. 1 (A-(i)) can be defined as $(V_1, E_1)$ where $V_1$={a,b,c} and $E_1$={(a,c), (c,b), (c,b), (b,a), (a,a)} whereas the undirected graph shown in Fig. 1(A-(ii)) can be defined as $(V_2, E_2)$, $V_2$={a,b,c} and $E_2$={{a,c}, {b,c}, {b,c}, {b,a}, {a,a}}. Note that, the order of vertices does not matter in the definition of the graph in Fig. 1(A-(ii)).

- An edge having identical *end* vertices is called a *self loop*. In the graph shown in Fig. 1(A-(i)) and Fig. 1(A-(ii)), (a, a) is *self loop*. If more than one edge is associated with a given pair of vertices then the edges are referred to as parallel edges. In graph A and B, $e_1$ and $e_2$ are parallel edges.
- A graph is said to be *connected* if there exists at least one path between every pair of vertices. Otherwise the graph is called *disconnected*. A *disconnected* graph consists of more than one *connected* subgraphs each of which is called a *connected component*. The graph shown in Fig. 1(B) consists of two connected components X and Y.
- A *walk* is a finite *alternating sequence* of vertices and edges beginning and ending with vertices, such that each edge is incident with vertices preceding and following it. In a walk, no edge appears more than once but vertices may appear more than once. In the graph shown in Fig. 1(B) a-$e_4$-c-$e_2$-b-$e_9$-p is a walk.
- A *closed walk* is a walk which begins and ends with same vertices. A closed walk in which no vertex appears more than once (except the initial and final vertex) is called cycle or circuit. In the graph shown in Fig. 1(A-(i)), a-$e_4$-c-$e_2$-b-$e_5$-a is a cycle.

### 2.4. *Bond* type (*b-type*) of nucleotide bases

Considering the hydrogen bonding nature of the nucleotide bases, two categories or ñbond-typesò, shortened as *b-type*, for nucleotide bases are defined. The first one is called *b-type* 1, containing A and U/T (2 hydrogen bonds) and the second one is called *b-type* 0 containing C and G (3 hydrogen bonds).

## 2.5. *Impact* of amino acids

Although *b-type* can demonstrate the hydrogen bonding nature of A, U/T, C and G (in RNA T is replaced by U), it is unable to uniquely identify four nucleotide bases. Information-theoretically, to uniquely recognize 4 nucleotide bases, $\log_2 4 = 2$ binary symbols are needed. In general, one could assign any permutation of the strings {00, 01, 10, 11} to the symbols A, U/T, C and G which generates 24 possible encodings. Given two strings over a set of symbols, the Hamming distance (Cover and Thomas, 2006) between them is the number of positions in which they differ. Thus, in some sense, it measures the similarity between the two strings ï the more the hamming distance, the less the similarity. It can be shown that the Hamming distance [*d*(.,.)] satisfies the definition of a metric, i.e., for any two strings *x* and *y*,

1. *d*(*x*,*y*) Ó0 (non-negativity),
2. *d*(*x*,*y*) = 0 if and only if x = y (coincidence),
3. *d*(*x*,*y*) = *d*(*y*,*x*) (symmetry), and
4. *d*(*x*,*y*) Ò*d*(*x*,*z*) + *d*(*y*,*z*) (triangle inequality).

In this context, any two strings in the above representation have a hamming distance of 1 or 2.

For the sake of simplicity in further analysis, we define the terminology of intra-*b-type* and inter-*b-type* distances as follows. For a particular encoding, intra-*b-type* distance is defined as the hamming distance between two strings representing nucleotides of same *b-type* value, e.g., for the encoding Aź 00, Tź 01, Cź 10, Gź 10, intra-*b-type* distance is *d(00, 01)= d(10, 11)=1*. It is to be noted that, *intra-b-type* distance for a particular encoding is always unique. On the other hand, inter-*b-type* distance is defined as the hamming distance between two strings representing nucleotides of different *b-type* values. For the above mentioned encoding of A, T, C and G, inter-*b-type* distance can be *d(00,10)=d(01,11)=1* or *d(00, 11)=d(01, 10)=2*. It should be noted that, unlike intra-*b-type* distance, inter-b-type distance for a particular encoding is not unique.

Thus, considering the similarity of *b-types*, there are two possibilities:
A) Either an intra-*b-type* distance of 1
or B) an intra-*b-type* distance of 2 .

As shown in Table S1, out of the 24 possible encodings, there are 16 encodings of category A (non shaded portion) and there are 8 encodings of category B (shaded portion). Note that the encodings of category A can have both 1 and 2 as inter-*b*-type distance, thereby leading to ambiguity. On the other hand, all the encodings of category B have inter-*b*-type distance of 1 only, leading to unique and uniform feature of both inter- and intra-*b*-type distances. For this reason, we stick to the encoding category B. Further, it has been found that the target biochemical properties can be linked to the combinatorial properties of encodings of B mentioned above and any one of the 8 possible encodings leads to the same results, as one would expect. Hence, without loss of generality, we fix the encoding Aź 00, Tź 11, Cź 01, Gź 10 in this study.

Now, any sequence XYZ of three nucleotide bases, where $X, Y, Z \in \{U, C, A, G\}$, representing an amino acid, can be considered as a sequence of 6 symbols $b_1b_2b_3b_4b_5b_6$,where ($b_1b_2$), ($b_3b_4$), ($b_5b_6$) are encodings of X, Y and Z respectively. Since the first two nucleotide bases mostly take the determining role towards coding an amino acid (e.g., Glycine be coded by GGU, GGC, GGA and GGG, the first two nucleotide bases are GG), the first 4 symbols $b_1b_2b_3b_4$ have been considered to define *Impact* of an amino acid as follows. The *Impact* is a pair ($I_1$, $I_2$), where $I_1$ and $I_2$ can be calculated as follows:

$$I_1 = ½\,[sym\,(b_1, b_3) + sym\,(b_2, b_4)],$$
$$I_2 = ½\,[sym\,(b_1, b_2) + sym\,(b_3, b_4)],$$

where
$$sym(b_i, b_j) = \begin{cases} 1, \text{ if } b_i \neq b_j \\ 0, \text{otherwise.} \end{cases}$$

To illustrate, for Glycine the encoding of the first two nucleotide bases, i.e., GG, is 1010; hence $I_1$= [*sym*(1,1) + *sym*(0,0)]/2=0; and $I_2$= [*sym* (1,0)+ *sym*(1,0)]/2=(1+1)/2=1. Thus, *Impact* of Glycine is (0, 1).

It should be noted that, the *Impact* of an amino acid remains invariant, even if the assignment of symbols to the nucleotide bases are changed, keeping intra-*b-type* distance 2. Calculated *Impact* of 20 amino acids is demonstrated in Table 1.

## 3. Results and Discussions

More than 50 years ago it was hypothesized by Chemistry Nobel Laureate Anfinsen C.B that the information regarding the secondary and tertiary structure of proteins is embedded in its primary amino acid sequence (Anfinsen 1973) and hence the functional aspect, as both the function and the structure are correlated. Utilizing the concept of óstructure function relationshipô of protein along with the proposition based on mathematical modeling and graph theory, here it is substantiated that the underlying impetus for recognition of DXCA exclusively by the protein PPCA is based on its local conformation which is guided by the ñfirst principle of foldingò.

### 3.1. Conformational aspect of interaction

From the crystal structure of PPCA (PDB ID- 1OS6) it is found that the amino acid residues present at $4^{th}$, $29^{th}$, $37^{th}$, $38^{th}$, $41^{st}$, $45^{th}$ and $50^{th}$ positions in the sequence participate in the interaction with four oxygen atoms of DXCA. As the homologs of PPCA cannot recognize DXCA, it will be worthy to identify whether the underlying hypothesis is implanted either in the primary sequence or in the secondary structure of these particular residues or simultaneously both.

### 3.2. Sequence alignment

Using the program BLAST (blastp) (Altschul et al. 1990) it has been observed that with respect to PPCA, the sequence identity of PPCB: ~77%, PPCC: ~65%, PPCD: ~62% and PPCE: ~58% which establishes that there exists at least ~25% of sequence differences among the homologs with respect to PPCA. More precisely if one looks into the interacting residues, it is found that isoleucine (I) in PPCA is replaced by methionine (M) in PPCB at $4^{th}$ position; lysine (K) (in PPCA) is replaced by valine (V) (in PPCB) at $29^{th}$ position and glycine (G) (in PPCA) is replaced by serine (S) (in PPCB) at $50^{th}$ position. In PPCC with respect to PPCA, mutation observed at the residue position $29^{th}$ and $37^{th}$, where lysine (K) (at $29^{th}$ position) is replaced by glycine (G) and lysine (K) (at $37^{th}$ position) is replaced by arginine (R) respectively. While in PPCD protein almost all the interacting residues are mutated. In PPCE mutation has been observed at residues $29^{th}$, $45^{th}$ and $50^{th}$ position. Comparison of primary amino acid sequence along with interacting residues of PPCA for DXCA and the other homologs of PPCA family is highlighted in Fig. 1.

### 3.3 Secondary Structure Analysis

Analysis of secondary structure of PPCA and its homologs (PPCB, PPCC, PPCD & PPCE) using DSSP program (Kabsch and Sander 1983) indicate that amino acid of each proteins at $4^{th}$ position have extended (E) conformation. At $29^{th}$ position amino acid of PPCA and PPCE protein adopt $3_{10}$ helical (G) conformations, while in PPCB amino acid at the same position is in Ŭ-helical (H) conformation. For protein PPCC and PPCD amino acid forms H-bonded turn (T) and bend (S) conformation respectively. From the aspect ofsecondary structure conformation of the binding residues, the most appealing scenariois found at $37^{th}$ and $38^{th}$ position which mainly involve in recognition of DXCA through H-bonding. Amino acids at $37^{th}$ and $38^{th}$ position of PPCA protein are within the Ŭ-helical (H) region; however, for the rest other homologs these are the part of PPII structure (Table 2) which is demonstrated by the superposition of the truncated secondary structures using the program Superpose (Maiti et al. 2004) (Fig S1). Amino acids at $41^{st}$ and $45^{th}$ position of all proteins are in coil and Ŭ-helical (H) conformation respectively. At $50^{th}$ position amino acid in PPCA, PPCB and PPCC are in H-bonded turn (T) conformation while rest (PPCD & PPCE) is in Ŭ-helical (H) conformation. Considering the interacting sites of PPCA for recognition of DXCA, one can conclude that there is a large deviation in the secondary structure of $37^{th}$ and $38^{th}$ residue (K & I) playing a major role in DXCA recognition with respect to other members of the PPCA family. Moreover, for other interacting positions, deviation in the respective primary sequence may not initiate the formation of hydrogen bond towards recognition of DXCA by the other homologs.

From the analysis of the sequence and the secondary structure of the proteins representing the PPCA family towards recognition of DXCA, it can be established that PPCA differs from its homologs at the interacting residues along with few others which may be correlated with its functionality as such information is embedded within the primary sequence. Moreover, it can be emphasized that for adopting a distinct conformation at a particular residue there would be a role of neighboring amino acids. Justification of exclusive interaction of PPCA with DXCA is validated through the graph theoretic approach in the subsequent part of this manuscript.

### 3.4 Based on combinatorial analysis

In order to justify the underlying hypothesis for the behavioral difference of PPCA with DXCA as concluded from the aspect of óstructure-function relationshipô, two graph theoretic models have been developed based on *b-type* of nucleotide bases and *Impact* of amino acids respectively.

### 3.4.1 Connectivity of graphs based on *b-type*

Each protein of PPCA family has been characterized by a ôbinding sequenceô which contains the interacting location wise ordered sequence of amino acid residues. Thus the binding sequence for PPCA is ñI (4th position) K (29th position) K (37th position) I (38th position) F (41st position) M (45th position) G (50th position)ò (Fig. 1).

For any given protein of PPCA family, we denote each interacting residue of the ñbinding sequenceò as $P_{ij}$ where *i* (=1, 2, 3, 4, 5, 6, 7) denotes the position of the interacting residues in the binding sequence and *j* (=1, 2, 3) denote the positions of nucleotide bases in the interacting residues (shown in Table 3). Corresponding to each member of PPCA protein family, we draw two separate directed graphs (Fig. S4), namely $G_0= (V_0, E_0)$ and $G_1= (V_1, E_1)$ based on b-type (1 or 0) values as follows:

$V_0= \{P_{ij}: \textit{b-type}\ (P_{ij}) = 0\}$ and

$E_0 = \{(P_{ij}, P_{(i+1)j'}): P_{ij} \in V_0, P_{(i+1)j'} \in V_0, |j - j'| \leq 1\}$;

$V_1= \{P_{ij}: \textit{b-type}\ (P_{ij}) = 1\}$ and

$E_1 = \{(P_{ij}, P_{(i+1)j'}) : P_{ij} \in V_1, P_{(i+1)j'} \in V_1, |j - j'| \leq 1\}$;

To illustrate, let us consider the construction of graph $G_0$ for PPCA. From the second column of PPCA (depicted in Table 3), it may be noted that, for L=4, i=1, *b-type* of $P_{13}$(= C) is 0; for L=29, i=2, *b-type* of $P_{23}$(=G) is 0. As a result, there is an edge between the nodes representing $P_{13}$ and $P_{23}$ in $G_0$. Similar logic holds for the rest of the graph as well as for $G_1$.

From the graph theoretic representation (Fig. S2) of different proteins of PPCA family, it can be observed that, in the graph of PPCA, each $G_0$ and $G_1$ consists of only one connected component. For itôs other homolog, more than one connected component exists in either $G_0$ or $G_1$. This demonstrates this differential nature of PPCA from itôs homolog.

*It should be noted that, altering the b-type values (subject to intra-b-type distance 2) also produces the same result.*

### 3.4.2 Characterization of interacting residues using a graph theoretic approach based on *Impact*

Although the graph theoretic approach based on *b-type* values of the nucleotide bases establish the characteristic uniqueness of PPCA from its other homologs, no explicit indication can be obtained regarding the underlying chemical phenomenon. So, at this juncture it would be worthy to justify whether this exceptional property can be explained further from the ñfirst principle of protein foldingò through developing a combinatorial model based on the *Impact* of amino acids. Table 3 represents the interacting location wise demonstration of amino acid residues for each protein of PPCA family in terms of Nucleotide codons (N.C.) (obtained from the óncbiô http://www.ncbi.nlm.nih.gov/) along with the *b-type* values of corresponding nucleotide bases in triplet form, binary representations (b.r.) of codons and *Impact* values.

An ñalternative representationò of ñbinding sequenceò, defined in the previous section, can be given in terms of *Impact* values ($I_{1i}$, $I_{2i}$), where *i*=1, 2, 3, 4, 5, 6, 7 represents the location of interacting residues in binding locations. Thus the alternative representation of PPCA is {(1, 0), (0, 0), (0, 0), (1, 0), (0, 0), (1, 0), (0, 1)}. Since there exists only five *Impact* values (depicted in Table 1) and seven ñbinding locationsò, repetition of *Impact* values in ñalternative representationò is obvious (pigeon-hole principle). From the ñalternative representationò, a directed graph G = {V, E}, can be drawn for each protein of PPCA family, where V = distinct pairs from $\{(I_{1i}, I_{2i})$: i=1, 2, 3, 4, 5, 6, 7} and E= $\{((I_{1i}, I_{2i}), (I_{1j}, I_{2j})): j=i+1\}$. It may be noted that although serine (S) is having two impact values that constructed graphs remain unchanged even if we take the second value of serine.

To illustrate, let us consider the following example: for both L=4, i=1 and L=29, i=2 ($I_{1i}$,$I_{2i}$) is (1, 0) and (0, 0) respectively (depicted in Table 3). So, an edge has been drawn between the nodes representing the nodes (1, 0) and (0, 0). On the other hand, ($I_{1i}$, $I_{2i}$) is (0, 0) for both L=29, i=2 and L=37, i=3. So, a self loop has been drawn on the node (0, 0), Parallel edges are not shown, as they do not signify any special property. In the graphical representation (depicted in Fig. 3), each vertex, representing an interacting amino acid has been shown in the form of $A_i$ ($C_i$ $L_i$) where $A_i$ indicates the amino acid residue, $C_i$ represents the characteristics feature of the amino acid according to Chou-Fasman scale (Chou and Fasman, 1974) [helix breaker (B), helix former (H) or helix indifferent(i)] and $L_i$ denotes the location of the amino acid in the respective protein. It can be observed that, the graphs for PPCB to PPCE contain a cycle of length 3 which is absent in the graph of PPCA as it contain a cycle of length 2. The graphical representations remain unchanged even if the codes assigned to the different nucleotide bases are altered, keeping intra-*b-type* hamming distance 2.

This observation establishes the characteristic uniqueness of PPCA from its other homologs, as described in previous sections.

It is also important to note that, in the graphs of PPCB to PPCE, the vertices corresponding to helix breaker (B) (e.g. Glycine) or helix indifferent (i) (e.g. Arginine, Serine) are part of the cycles for length 3 while in PPCA the vertex representing the helix breaker Glycine (at $50^{th}$ position) does not belong to the cycle of length 2. *This observation focuses on the helix formation of PPCA which is essential for recognition of DXCA, making PPCA a different entity from its homologs.*

## Conclusion

Difference in behavior of PPCA towards recognition of DXCA in comparison to its other homologs in $C_7$ family clearly results from the difference in the primary sequence and its secondary structure at the ‵interacting regionô which is actually guided by their genomic sequence as converged from the result of structural biology and discrete mathematics. It can be emphasized that the function of a protein is in fact a synchronized effect of the overall sequence and structure of the binding sequence comprised of interacting amino acids along with their neighboring residues. Combinatorial analysis of sequences and graph theoretic models thus imply the harmony among the interacting residues enabling to understand the functional network of both the protein and the gene.

## Acknowledgement

The authors acknowledge BIF-DBT centre WBUT for computational facility, Indian Statistical Instituteǿs internal project fund and acknowledge Dr. Pokkuluri, Phani R (Argonne Lab) for discussing the problem.

## REFERENCES

Anfinsen CB. (1973) Principles that Govern the Folding of Protein Chains. *Science*. 181: 223ï 230.

Alberts B, Johnson A, Lewis J, Raff M, Roberts K, Walter P (2002). Molecular Biology of the Cell (Fourth ed.). New York: Garland Science. ISBN 978-0-8153-3218-3.

Altschul S, Gish W, Miller W, Myers E , Lipman D (1990). Basic local alignment search tool. *Journal of Molecular Biology* **215** (3): 403ï 410. doi:10.1016/S0022-2836(05)80360-2.

Bond DR, Lovley DR. (2003) Electricity production by Geobactersulfurreducens attached to electrodes. Appl. Environ. Microbiol.69:1548-1555.

Burgess R., Deutscher M. (1990) Guide to Protein Purification. Methods in Enzymology. Academic Press, San Diego. Vol. 182

Chou P. Y., Fasman G. D. (1974) Prediction of Protein Conformation Biochemistry,13: 222-245.

Caccavo F, Lonergan DJ, Lovley DR, Davis M, Stolz JF, McInerney MJ. (1994) Geobactersulfurreducens sp. nov., a hydrogen- and acetate-oxidizing dissimilatory metal-reducing microorganism. Appl. Environ. Microbiol. 60:3752-3759.

CoverThomas M, ThomasJoy A, (2006) Elements of Information Theory, Wiley, 2nd Ed.

Deo N. Graph Theory with Applications to Engineering and Computer Science (2003),PHI, $25^{th}$ edition.

Jacobs D. J.,Rader A. J.,Kuhn L. A., Thorpe M. F. (2001) Protein Flexibility Predictions Using Graph Theory. PROTEINS: Structure, Function, and Genetics. 44:150ï 165.

Kabsch W, Sander C. (1983) Dictionary of protein secondary structure: pattern recognition of hydrogen-bonded and geometrical featuresò. Biopolymers . 22: 2577-2637.

Lloyd JR, Leang C, Hodges Myerson AL, Coppi MV, Cuifo S, Methe M, Sandler SJ, Lovley DR. (2003) Biochemical and genetic characterization of PpcA, a periplasmic c-type cytochrome in Geobactersulfurreducens, Biochem. J. 369: 153ï 161.

Larkin MA, Blackshields G, Brown NP, Chenna R, McGettigan PA, McWilliam H, Valentin F, Wallace IM, Wilm A, Lopez R, Thompson JD, Gibson TJ, Higgins DG. ClustalW and ClustalX version 2. (2007) Bioinformatics . 23 (21): 2947ï 2948.

Leavitt, Sarah A. (June 2010). "Deciphering the Genetic Code: Marshall Nirenberg". Office of NIH History.

Maiti R, Domselaar GHV, Zhang H, and Wishart DS. (2004) SuperPose: a simple server for sophisticated structural superposition, Nucleic Acids Res. 1; 32 (Web Server issue): W590W594.

Morgado L, Paixão VB, Schiffer M, Pokkuluri PR, Bruix M, Salgueiro CA. (2012) Revealing the structural origin of the redox-Bohr effect: the first solution structure of a cytochrome from Geobactersulfurreducens. Biochem J. 441:179-187.

Pokkuluri PR., Londer YY, Duke NEC, Pessanha M, Yang X, Orshonsky V, Orshonsky L, Erickson J, Zagyanskiy Y, Schiffer M. Structure of a novel dodecaheme cytochrome c from Geobacter.sulfurreducens reveals an extended 12 nm proteinwith interacting hemes. Journal of structural biology. 174: 223-233.

Pokkuluri PR, Londer YY, Duke NEC, Long WC, Schiffer M. (2004) Family of cytochrome c7-type proteins from Geobactersulfurreducens: Structure of one cytochrome $c_7$ at 1.45 Å resolutions. Biochemistry 43: 849-859.

Pokkuluri PR, Londer YY, Yang X, Duke NEC, Erickson J, Orshonsky V, Johnson G, and Schiffer M. (2010) Structural characterization of a family of cytochromes c7 involved in Fe(III) respiration by Geobactersulfurreducens. Biochim. Biophys. Acta Bioenergetics 1797: 222-232.

Shelobolina ES, Coppi MV, Korenevsky AA, Didonato LN, Sullivan SA, Konishi H, Xu H, Leang C, Butler JE, Kim BC, Lovley DR. (2007) Importance of c-type cytochromes for U(VI) reduction by Geobactersulfurreducens. BMC Microbiol. 7: 16.

Shi L, Squier TC, Zachara JM, Fredrickson JK. (2007) Respiration of metal (hydr) oxides by Shewanella and Geobacter: a key role for multi haem c-type cytochromes. Mol. Microbiol. 65:12ï 20.

Slack, J.M.W. Genes-A Very Short Introduction. Oxford University Press 2014

# Appendix:

## 1. Figures

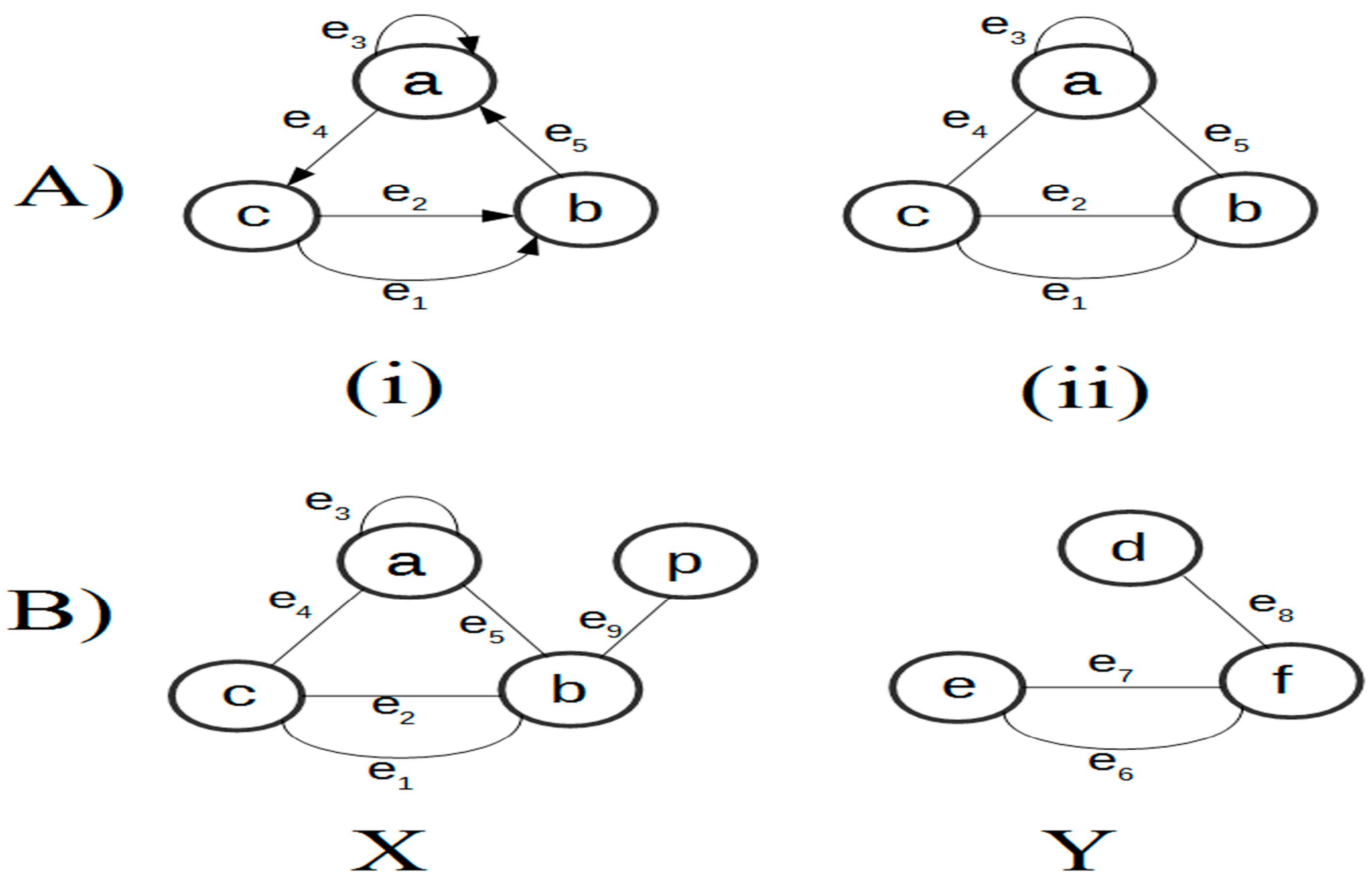

**Fig. 1: G**raph theoretic terminologies: (A-i) Directed graph. (A-ii) Undirected graph. (B) Demonstration of the concept of connected components.

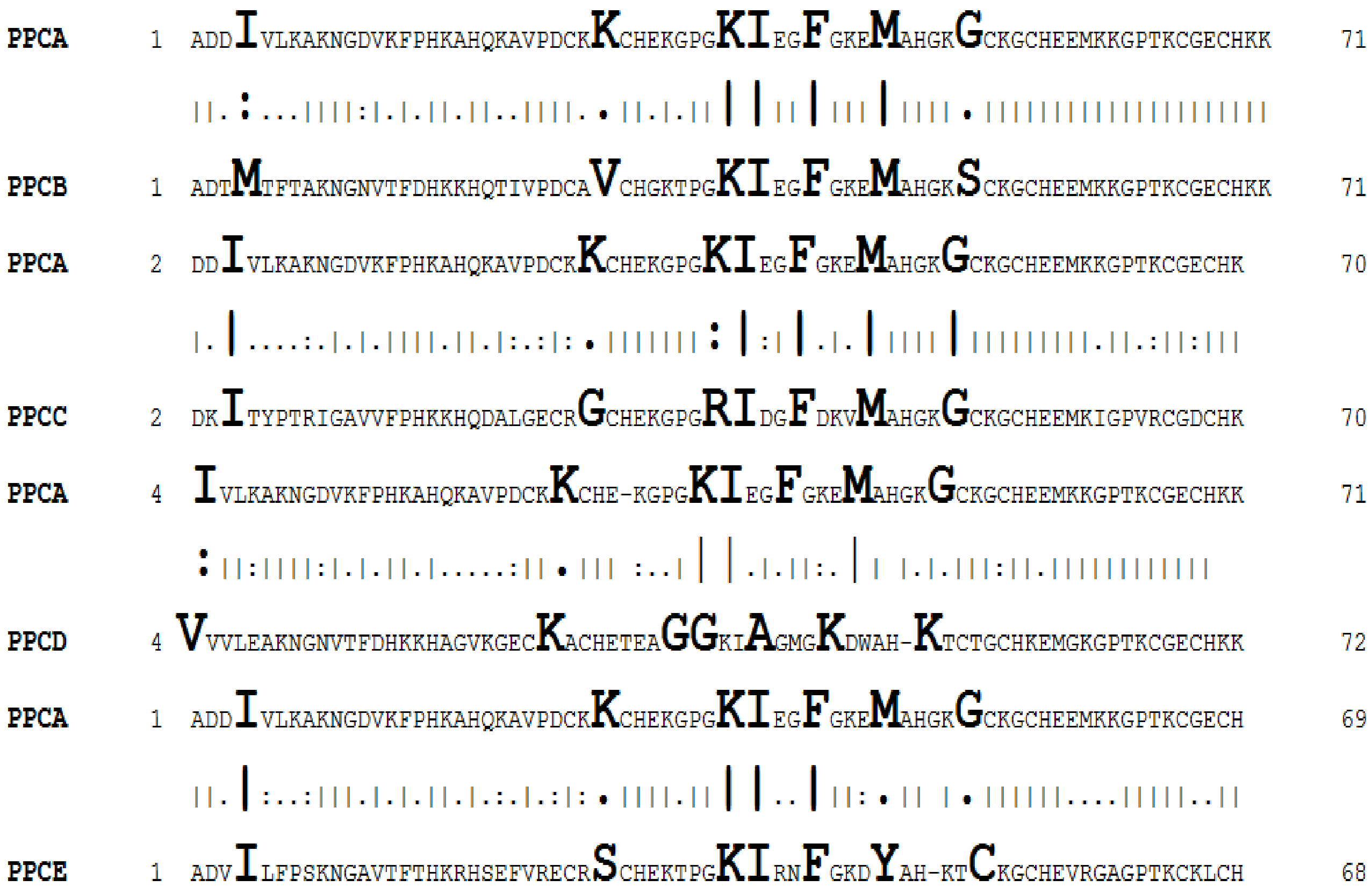

**Fig. 2:** Pair wise sequence alignment of the primary sequences of PPCA and its homologs (with respect to PPCA) obtained from cytochrome $C_7$ found in *Geobacter sulfurreducens* using Clustal W.

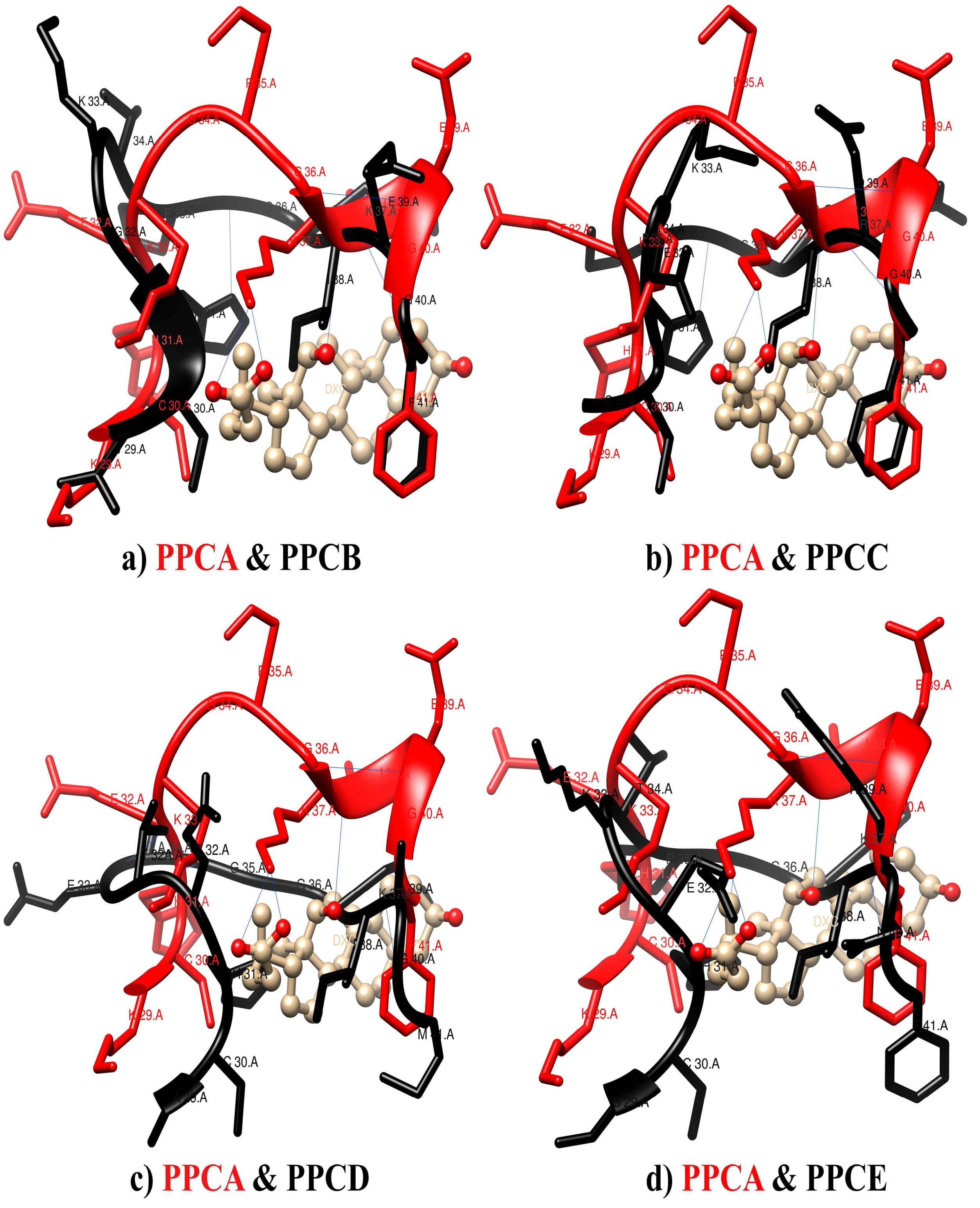

**Fig. 3:** Cartoon representation of overlay of secondary structural elements (from position 29 to 41) between PPCA (shown in red color) and its different homologs (shown in black color) obtained from cytochrome $C_7$ using Superpose server towards understanding the exclusive interaction between DXCA (shown in ball & stick model) and PPCA.

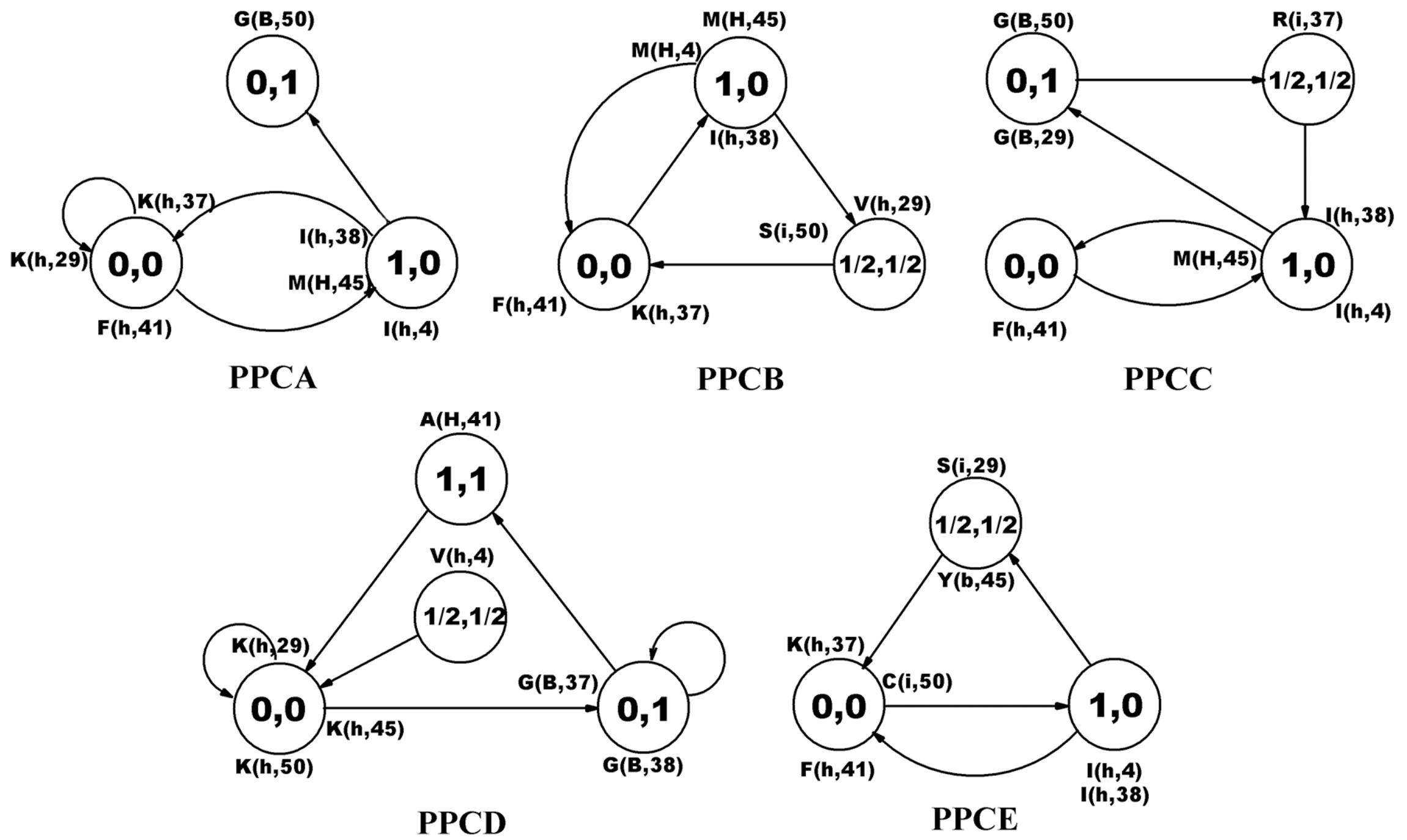

**Fig. 4 :**Graphical representation of each interacting amino acid based on the ñImpactò of the residues of ñbinding sequencesò of PPCA and the residues at similar position of the homologs of PPCA showing that there are three cycle of length 1, 2 and 3 in the graphs representing, whereas PPCB to PPCE contain two cycles of length 1 and 3 or 2 and 3.

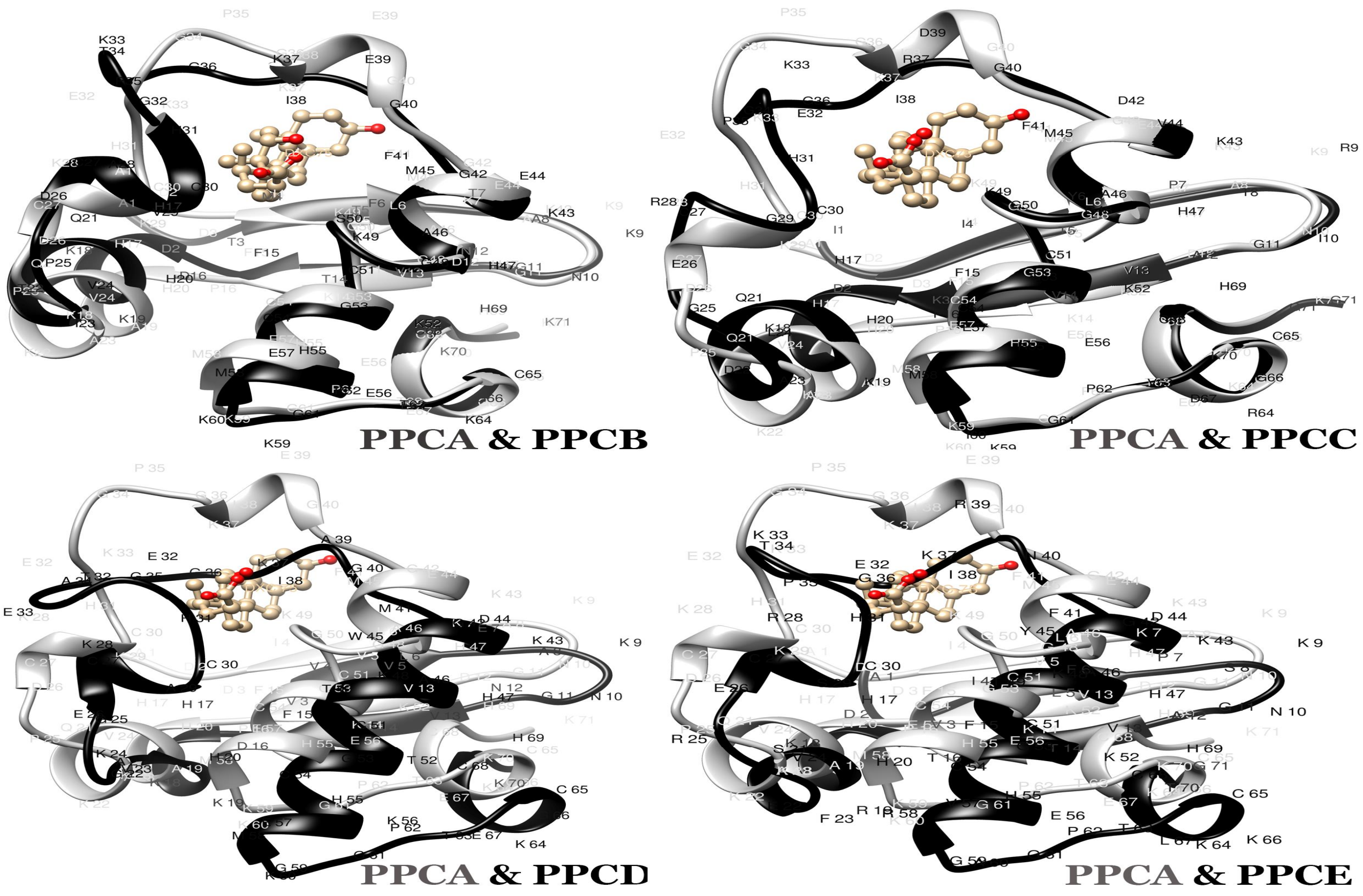

**Fig S1:** Cartoon representation of overlay of secondary structural elements of full protein between PPCA (shown in red color) and its different homolog.

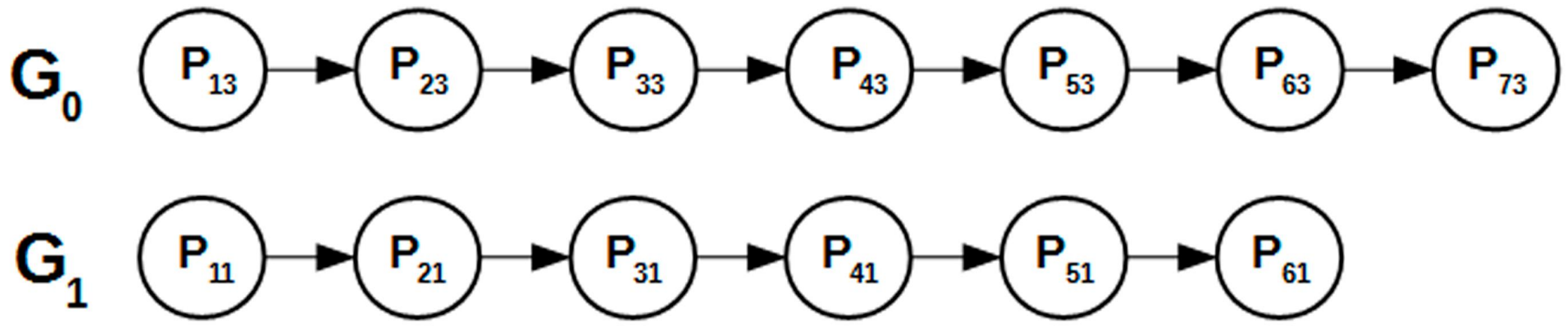

(A) Graphical representation of PPCA based on *b-type*: $G_0$ : *b-type*=0; $G_1$ : *b-type*=1

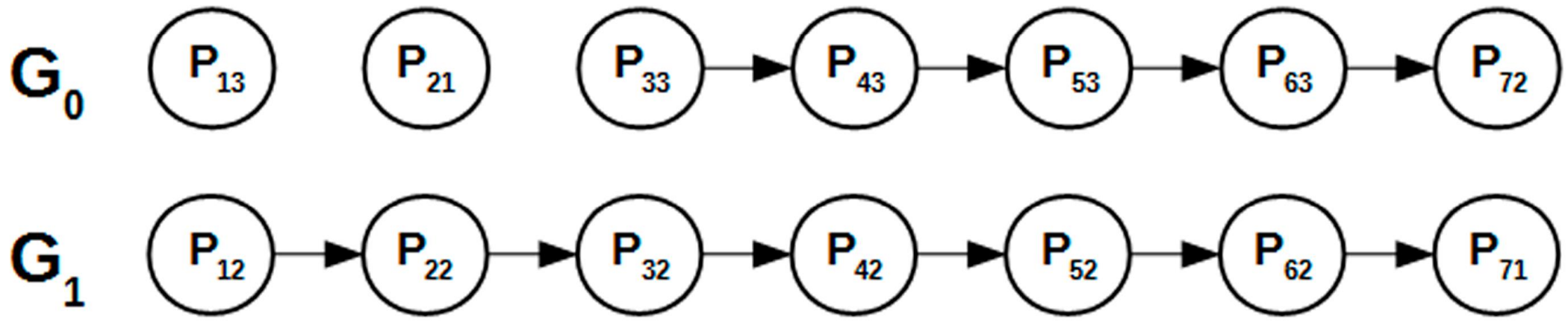

(B) Graphical representation of PPCB based on *b-type*: $G_0$ : *b-type*=0; $G_1$ : *b-type*=1

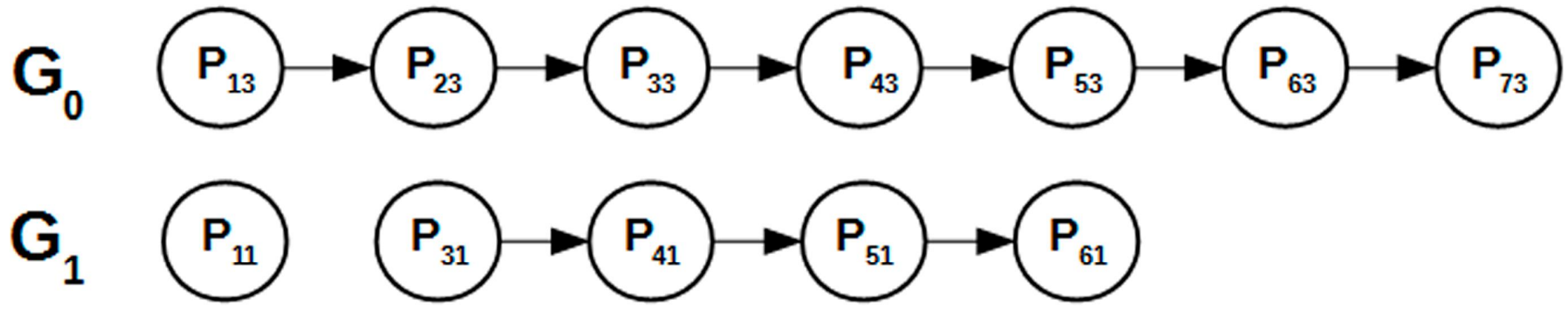

(C) Graphical representation of PPCC based on *b-type*: $G_0$ : *b-type*=0; $G_1$ : *b-type*=1

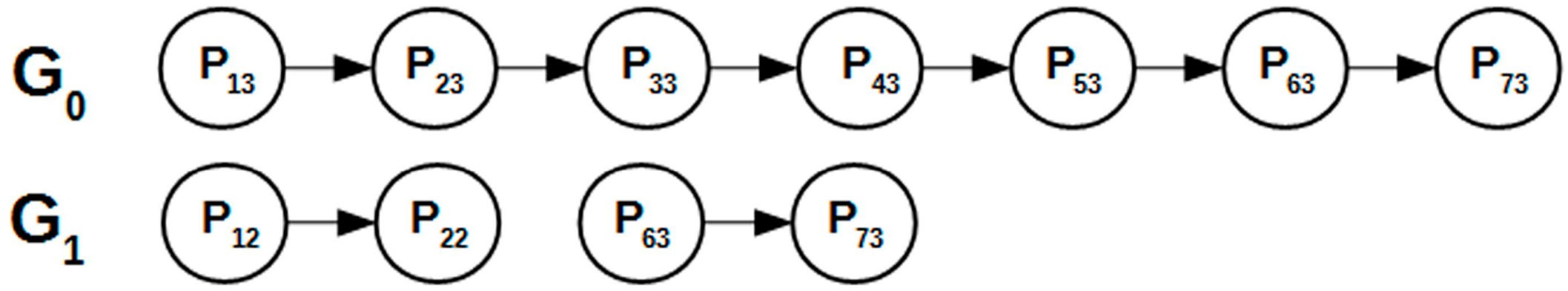

(D) Graphical representation of PPCD based on *b-type*: $G_0$ : *b-type*=0; $G_1$ : *b-type*=1

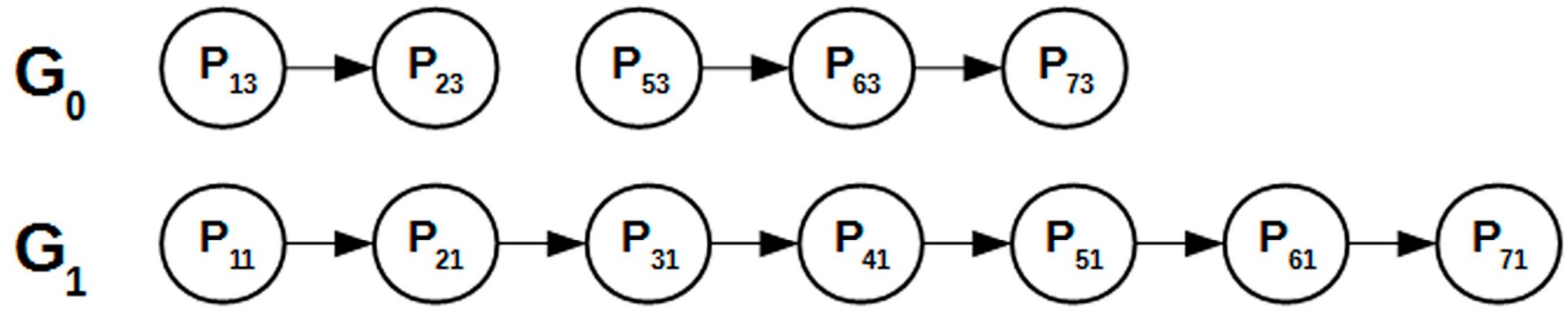

(E) Graphical representation of PPCE based on *b-type*: $G_0$ : *b-type*=0; $G_1$ : *b-type*=1

**Fig. S2:** Graphical representations of each PPC family protein based on *b-types*, showing that for PPCA both $G_0$ and $G_1$ contain only one connected component, while for others more than one connected components exist either in $G_0$ or in $G_1$ or in both.

## 2. Tables

| Amino acid | Impact | Amino acid | Impact |
|---|---|---|---|
| Ala(A) | (1,1) | Met(M) | (1, 0) |
| Cys(C) | (1/2,1/2) | Asn(N) | (0, 0) |
| Asp(D) | (1/2,1/2) | Pro(P) | (0, 1) |
| Glu(E) | (1/2,1/2) | Gln(Q) | (1/2,1/2) |
| Phe(F) | (0, 0) | Arg(R) | (1/2,1/2) |
| Gly(G) | (0,1) | Ser(S) | (1/2,1/2) |
| His(H) | (1/2,1/2) | Thr(T) | (1/2,1/2) |
| Lys(K) | (0, 0) | Val(V) | (1/2,1/2) |
| Leu(L) | (1/2,1/2) or (0,0) | Trp(W) | (1/2,1/2) |
| Ile(I) | (1, 0) | Tyr(Y) | (1, 0) |

**Table 1: Calculated "impact" values of 20 amino acids**

| | PPCA | | | PPCB | | | PPCC | | | PPCD | | | PPCE | | |
|---|---|---|---|---|---|---|---|---|---|---|---|---|---|---|---|
| Position | Pattern | PHI | PSI | Pattern | PHI | PSI | Pattern | PHI | PSI | Pattern | PHI | PSI | Pattern | PHI | PSI |
| 4 | E | -120.2 | 148.9 | E | -100.8 | 147.6 | E | -124.7 | 117 | E | -92.4 | 109.3 | E | -115.6 | 133.5 |
| 29 | G | -64.3 | -27.7 | H | -52.8 | -41.9 | T | -58.9 | -52.5 | S | -109.6 | 153.6 | G | -65.8 | -19.2 |
| 37 | H | -75.8 | -23.9 | | -63.5 | 136.5 | | -60.1 | 145.6 | | -116.7 | 101.1 | | -58.4 | 140.3 |
| 38 | H | -72.5 | -46.8 | | -89 | 99.9 | | -103.9 | 110.6 | T | -57.6 | -34.2 | | -99.9 | 112.6 |
| 41 | | -69.3 | 150.7 | | -56.8 | 137.2 | | -70.6 | 118.3 | S | 165.5 | -170.8 | | -64.5 | 139.2 |
| 45 | H | -64.4 | -40.6 | H | -67.6 | -44 | H | -72.2 | -44.5 | H | -67.4 | -44.1 | H | -60.5 | -45.1 |
| 50 | T | -140.2 | -76.4 | T | -136.8 | -77.9 | T | -143.8 | -79.6 | H | -68.5 | -39.6 | H | -64.9 | -62.1 |

**Table 2: Representation of secondary structure obtained from DSSP along with backbone dihedral (ϕ, ψ) for the residues of PPCA participating for recognition of DXCA and the residues at similar position of PPCA homologs.**

| b-type | Nucleotide base | Intra *b-type* distance 1 | | | | | | | | | | | | | | | | Intra *b-type* distance 2 | | | | | | | |
|---|---|---|---|---|---|---|---|---|---|---|---|---|---|---|---|---|---|---|---|---|---|---|---|---|---|
| 1 | A | 00 | 00 | 01 | 01 | 00 | 00 | 10 | 10 | 10 | 11 | 10 | 11 | 01 | 11 | 01 | 11 | 00 | 00 | 11 | 11 | 01 | 01 | 10 | 10 |
| | T | 01 | 01 | 00 | 00 | 10 | 10 | 00 | 00 | 11 | 10 | 11 | 10 | 11 | 01 | 11 | 01 | 11 | 11 | 00 | 00 | 10 | 10 | 01 | 01 |
| 0 | C | 10 | 11 | 10 | 11 | 01 | 11 | 01 | 11 | 00 | 00 | 01 | 01 | 00 | 00 | 10 | 10 | 01 | 10 | 01 | 10 | 00 | 11 | 00 | 11 |
| | G | 11 | 10 | 11 | 10 | 11 | 01 | 11 | 01 | 01 | 01 | 00 | 00 | 10 | 10 | 00 | 00 | 10 | 01 | 10 | 01 | 11 | 00 | 11 | 00 |
| | Category A: Inter *b-type* distance is either 1 or 2 | | | | | | | | | | | | | | | | | Category B: Inter *b-type* distance is always 1 | | | | | | | |

**Table S1: 24 possible encodings**

| Location of the interacting residues in protein (L) and in binding sequence (*i*) | PPCA | | | | PPCB | | | | PPCC | | | | PPCD | | | | PPCE | | | |
|---|---|---|---|---|---|---|---|---|---|---|---|---|---|---|---|---|---|---|---|---|
| | Residue | N.C. (*b-type*) *j* 1 2 3 | b.r. | Impact ($I_{1i}$, $I_{2i}$) | Residue | N.C. (*b-type*) *j* 1 2 3 | b.r. | Impact ($I_{1i}$, $I_{2i}$) | Residue | N.C. (*b-type*) *j* 1 2 3 | b.r. | Impact ($I_{1i}$, $I_{2i}$) | Residue | N.C. (*b-type*) *j* 1 2 3 | b.r. | Impact ($I_{1i}$, $I_{2i}$) | Residue | N.C. (*b-type*) *j* 1 2 3 | b.r. | Impact ($I_{1i}$, $I_{2i}$) |
| L=4, *i*=1 | I | ATC (1, 1, 0) | 001101 | (1, 0) | M | ATG (1, 1, 0) | 001110 | (1,0) | I | ATC (1, 1, 0) | 001101 | (1,0) | V | GTG (0, 1, 0) | 101101 | (1/2, 1/2) | I | ATC (1, 1, 0) | 001101 | (1, 0) |
| L=29, *i*=2 | K | AAG (1, 1, 0) | 000010 | (0, 0) | V | GTA (0, 1, 1) | 101100 | (1/2, 1/2) | G | GGC (0, 0, 0) | 101001 | (0, 1) | K | AAG (1, 1, 0) | 000010 | (0,0) | S | AGC (1, 0, 0) | 001001 | (1/2, 1/2) |
| L=37, *i*=3 | K | AAG (1, 1, 0) | 000010 | (0, 0) | K | AAG (1, 1, 0) | 000010 | (0,0) | R | AGG (1, 0, 0) | 001010 | (1/2, 1/2) | G | GGG (0, 0, 0) | 101010 | (0, 1) | K | AAA (1, 1, 1) | 000000 | (0,0) |
| L=38, *i*=4 | I | ATC (1, 1, 0) | 001101 | (1, 0) | I | ATC (1, 1, 0) | 001101 | (1,0) | I | ATC (1, 1, 0) | 001101 | (1, 0) | G | GGG (0, 0, 0) | 101010 | (0, 1) | I | ATA (1, 1, 1) | 001100 | (1, 0) |
| L=41, *i*=5 | F | TTC (1, 1, 0) | 111101 | (0, 0) | F | TTC (1, 1, 0) | 111101 | (0,0) | F | TTC (1, 1, 0) | 111101 | (0, 0) | A | GCC (0, 0, 0) | 100101 | (1, 1) | F | TTC (1, 1, 0) | 111101 | (0, 0) |
| L=45, *i*=6 | M | ATG (1, 1, 0) | 001110 | (1, 0) | M | ATG (1, 1, 0) | 001110 | (1,0) | M | ATG (1, 1, 0) | 001101 | (1, 0) | K | AAG (1, 1, 0) | 000010 | (0, 0) | Y | TAC (1, 1, 0) | 110001 | (1,0) |
| L=50, *i*=7 | G | GGC (0, 0, 0) | 101001 | (0, 1) | S | AGT (1, 0, 1) | 001011 | (1/2, 1/2) | G | GGC (0, 0, 0) | 101001 | (0,1) | K | AAG (1, 1, 0) | 000010 | (0, 0) | C | TGC (1, 0, 0) | 111001 | (1/2, 1/2) |

**Table 3: Binary representations (b.r.) of nucleotide codon (N.C.) of the interacting residues of PPCA obtained from NCBI and the residues at similar position of PPCA homologs along with the calculated *b-type* (in parenthesis below the corresponding N.C.) and *impact* of the amino acids.**